\documentclass[preprint,proceedings]{rmaa}
\usepackage{paralist}

\SetYear{2003}
\SetConfTitle{Compact Binaries in the Galaxy and Beyond}

\title{Observational studies of Cataclysmic Variable evolution: \\
 Of samples, biases and surprises} 

\author{B.T. G\"ansicke\altaffilmark{1}}

\altaffiltext{1}{Department of Physics, University of Warwick, Coventry CV4
  7AL, UK.}

\shortauthor{G\"ansicke}
\shorttitle{Observational studies of CV evolution}

\fulladdresses{\item Department of Physics, University of Warwick, Coventry CV4
  7AL, UK (Boris.Gaensicke@warwick.ac.uk).}

\listofauthors{B. T. G\"ansicke}
\indexauthor{G\"ansicke, B. T.}

\abstract{I present brief status reports on three large observational
projects that are designed to test our current understanding of the
evolution of cataclysmic variables (CVs): The spectroscopic selection
of new CVs in the Hamburg Quasar Survey, the search for pre-CVs based
on Sloan colours and UK Schmidt/6dF multiobject spectroscopy, and the
identification of CVs that descended from supersoft X-ray binaries
using a \textit{HST}/STIS far-ultraviolet spectroscopic survey.}

\resumen{I present brief status reports on three large observational
projects that are designed to test our current understanding of the
evolution of cataclysmic variables (CVs): The spectroscopic selection
of new CVs in the Hamburg Quasar Survey, the search for pre-CVs based
on Sloan colours and UK Schmidt/6dF multiobject spectroscopy, and the
identification of CVs that descended from supersoft X-ray binaries
using a \textit{HST}/STIS far-ultraviolet spectroscopic survey.}

\addkeyword{Stars: Cataclysmic Variables}
\addkeyword{Stars: Evolution}

\begin{document}
\maketitle

\section{Introduction}
The essential idea of the standard model of cataclysmic variable (CV)
evolution (disrupted magnetic braking, King 1988) is that CVs evolves
towards shorter periods due to a combination of angular momentum
losses: magnetic braking (dominating in systems with orbital periods
$P_{\rm orb}\ga 3$\,h) and the less efficient gravitational radiation
(dominating in systems with orbital periods $P_{\rm orb}\la
2$\,h). The standard paradigm of CV evolution successfully explains
the 2--3\,h gap in the observed CV period distribution. However, most
other predictions made by this model are in strong contrast with the
properties of the known CV population.  Recently, a number of
far-reaching modifications for the standard scenario have been
proposed (see Spruit \& Taam 2001, King \& Schenker 2002, Schenker \&
King 2002 and Andronov et al. 2003). Unfortunately, none of them has
been completely successful in tuning the predictions so that they
fully agree with the observations. It is apparent that the disturbing
disagreements between theory and observations have a common
denominator: the possible impact of selection effects on the currently
known population of CVs. In order to quantitatively test any theory of
CV evolution it is necessary to establish a large and unbiased sample
of CVs as well as of their progenitors, (pre-CVs, detached white
dwarf/late type main sequence stars). Such an observational data base
will also serve for future improvements of the theory of CV
evolution.

\section{CVs from the HQS}
Most CVs have been discovered through one of the following three
channels: variability, blue colour, or X-ray emission. We are
currently pursuing a large search for CVs based primarily on their
\textit{spectroscopic} properties, using the Hamburg Quasar Survey
(HQS; Hagen et al. 1995) as target sample (G\"ansicke et al. 2002).
The HQS provides an efficient means of discovering CVs that are so far
under-represented in the currently known sample of CVs, i.e. weak
X-ray emitters, not particularly blue objects, and CVs that show
variability with low amplitudes or long recurrence times.  A careful
study of the CV discovering efficency of the HQS concluded that this
survey is especially sensitive to short period systems, such as
SU\,UMa dwarf novae (G\"ansicke et al. 2002).

So far, we have discovered 53 new CVs, including a number of
fascinating systems (Araujo-Betancor et al. 2004, G\"ansicke et
al. 2004), and doubling the number of known CVs in the sky
area/magnitude range covered by the survey, and measured orbital
periods for 29 of these systems (Fig.\,1).  The simple fact that we do
not detect a large number of new short orbital period systems implies
that either they do not exist in the quantity predicted by the
theories, or that they look very different from the short period CVs
we know (e.g. have weak emission lines).  Another striking feature of
the HQS CV population is the large number systems in the 3--4\,h
orbital period range, most of which (at least 6) are new SW\,Sex stars
(Rodr\'\i guez-Gil et al. 2004).  Whereas SW\,Sex stars initially
appeared to be the ``freaks" among the CVs, they now turn out to be a
significant sub-class, and understanding their relation to the global
population of CVs is likely to be closely related to a major
improvement of our understanding of CV evolution as a whole.

\begin{figure}[!t]
\includegraphics[angle=-90,width=\columnwidth]{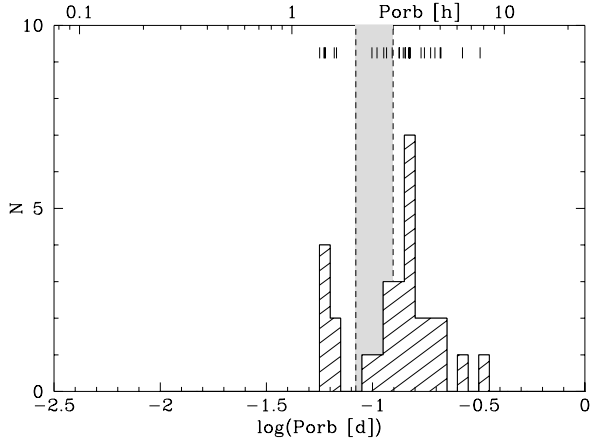} 
\caption{The period distribution of 29 new CVs from the HQS. An
  additional 24 HQS CVs are awaiting their period measurement. Shown
  in gray is the $2-3$\,h period gap. Note the concentration of
  systems in the $3-4$\,h period range.}
\end{figure}

\section{Pre-CVs from Sloan/UKST}
Whereas far more than 1000 CVs are listed by Downes et al. (2001),
only a shocking small number of pre-CVs is currently known. Schreiber
\& G\"ansicke (2003) have analysed in detail the properties of all
(30) pre-CVs for which both the orbital period and the white dwarf
temperature have been measured, and showed that this population is
heavily biased towards binaries containing rather hot (young) white
dwarfs ($T_\mathrm{wd}\ga15\,000$\,K) and low-mass/late-type
($M_\mathrm{sec}\la0.4M_\odot$) secondary stars. An important
consequence of this bias is that we currently know only \textit{a
single} progenitor for CVs with periods $\ga4$\,h:
V471\,Tau. Schreiber \& G\"ansicke also showed that this bias is a
natural result of the way that most pre-CVs were discovered~--~as a
white dwarf in the first place, with some evidence (weak emission
lines, eclipses, or ellipsoidal modulation) for a faint companion
cropping up later. Among the exceptions is V471\,Tau, which was indeed
discovered as a spectroscopic binary. Schreiber \& G\"ansicke conclude
that there ought to be a large population of pre-CVs containing cold
(old) and/or early-type secondary stars.

We have initiated a large search for pre-CVs with the aim to produce a
large and (largely) unbiased sample these systems. We select our
initial target list from Sloan Digital Sky Survey Data Release\,1
(Abazajian et al. 2003) by applying the following colour cuts:
$u-g<0.8$, $g-r<1.2$, and redder than the main-sequence.  A pilot
spectroscopic survey was carried out in October 2003 with the 6dF on
the 1.2\,m UK Schmidt Telescope, covering an area of $\simeq250\deg^2$
with  $g\la18.5$. In this run we identified 55 new white
dwarf/red dwarf binaries, a large fraction of them containing cold
white dwarfs (Fig.\,2).

An independent search for white dwarf/red dwarf binaries was carried
out by Raymond et al. (2003), also selecting targets on the base of
Sloan colours, but using the Sloan spectrograph for the identification
spectroscopy. The different colour selection employed by Raymond et
al. leads to a higher fraction of hot white dwarfs in their sample. 

At the current stage the orbital periods of the systems identified by
both projects are yet undetermined, and time-resolved follow-up
spectroscopy/photometry is necessary to distinguish the genuine pre-CVs
from wide binaries. Nevertheless, the results obtained so far clearly
demonstrate that combining multi-colour target selection with
multi-object identification spectroscopy will boost the number of known
pre-CVs over the next few years by a large factor.

\begin{figure*}
\includegraphics[angle=-90,width=\columnwidth]{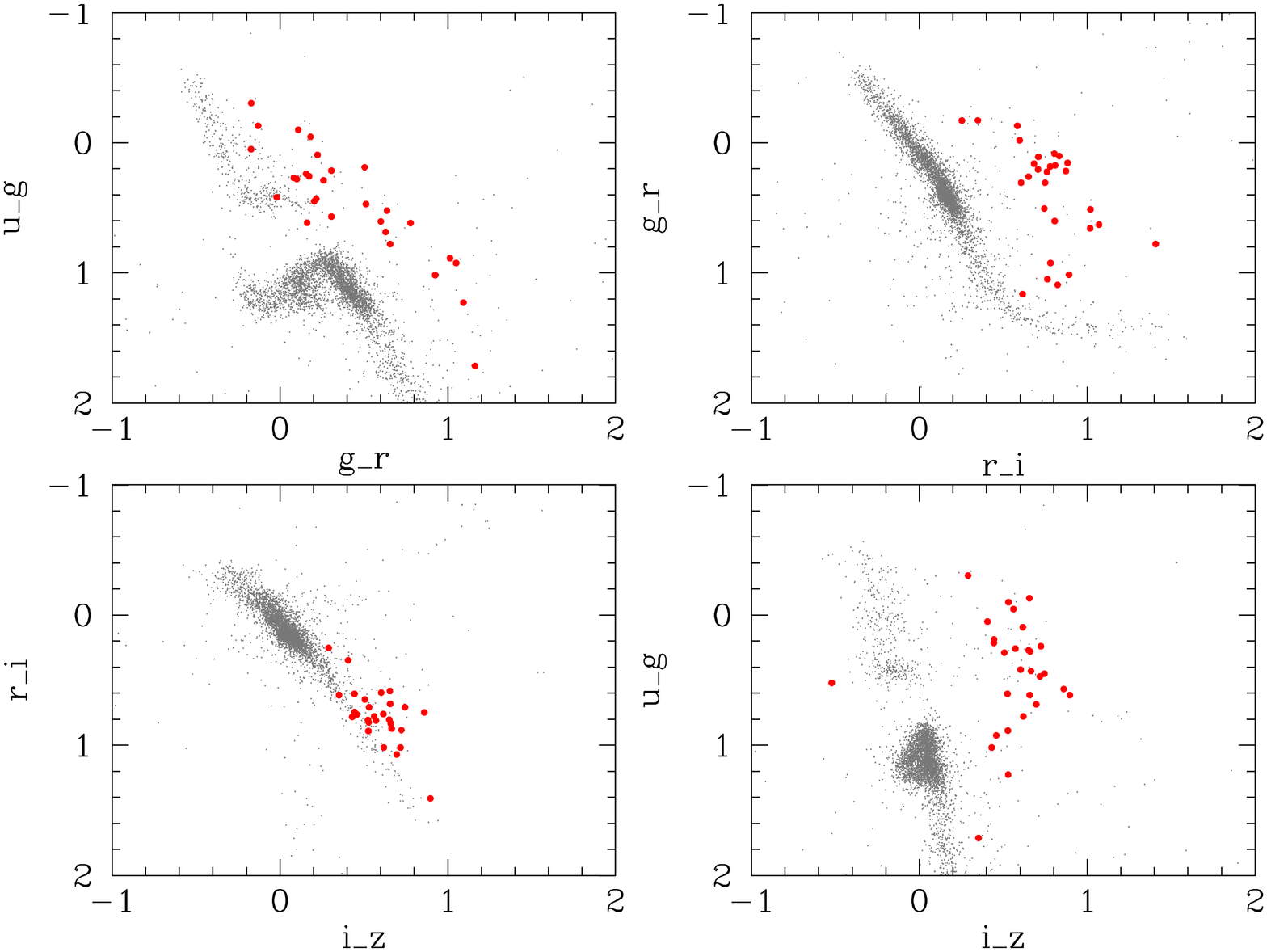}
\includegraphics[angle=-90,width=\columnwidth]{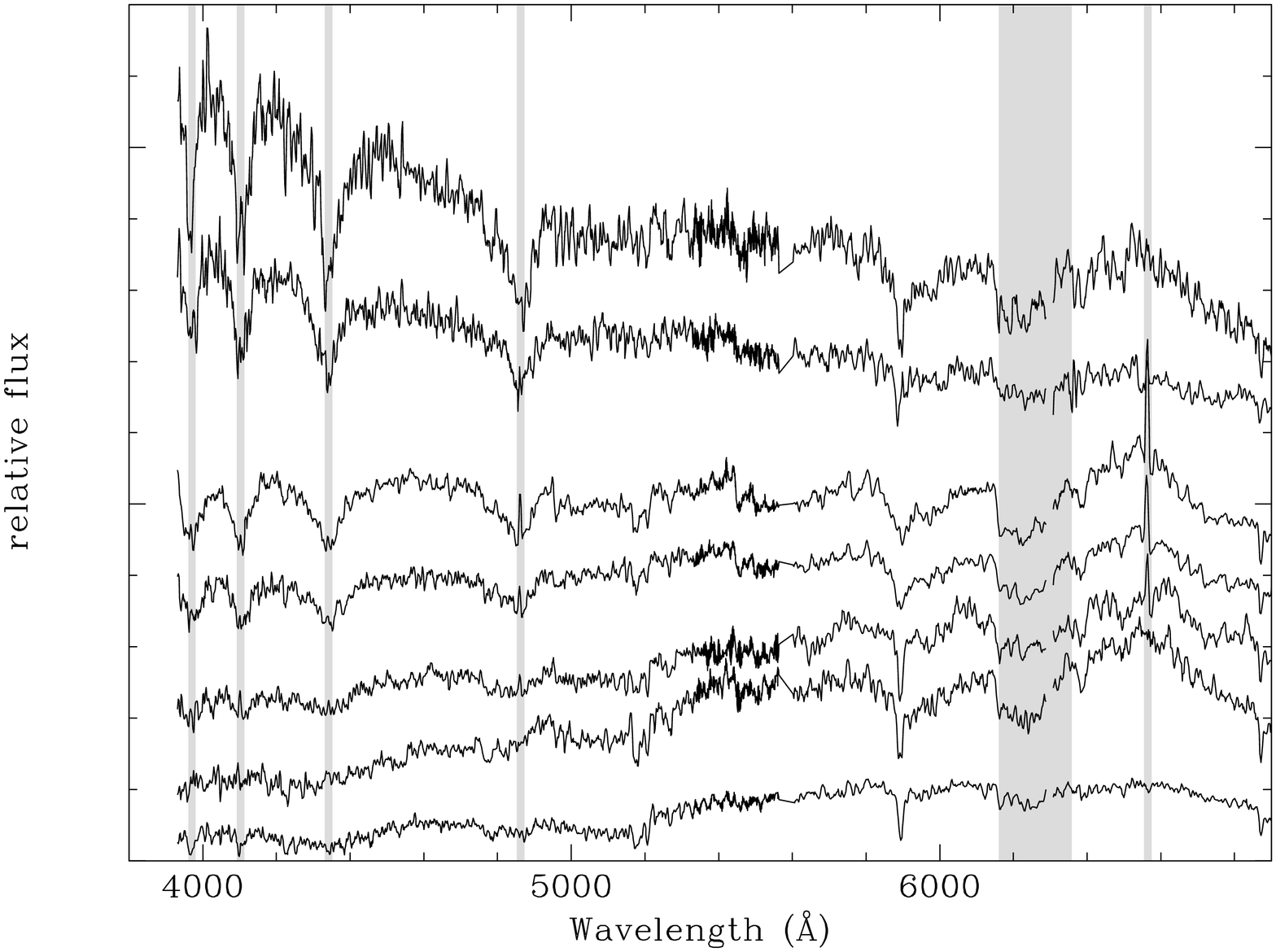}

\caption{\textit{Left panel:} Colour-colour diagrams of the white
 dwarf/red dwarf binaries identified from UKST/6dF observations. The
 small gray points represent the location of the main sequence and
 white dwarf population. \textit{Right panel:} Sample spectra of the
 UKST/6dF white dwarf/red dwarf binaries. The wavelength of the Balmer
 lines and the most prominent TiO absorption are shown in gray. The
 spectra are sorted by decreasing white dwarf contribution.}
\end{figure*}

\section{Post-supersoft CVs\,=\,failed SN\,Ia?}
Supersoft X-ray binaries are one likely channel for producing
supernovae of type Ia (e.g. Langer et al. 2000). In these systems,
contrary to ``normal" CVs, the donor star is more massive than the
white dwarf, and mass transfer occurs on the thermal time scale of the
donor star (Schenker et al. 2002). If the white dwarf fails to grow
over the Chandrasekhar limit, the mass ratio will eventually change
over ($M_\mathrm{wd}>M_\mathrm{donor}$), and the system will appear as
an apparently ``normal" CV~--~with the difference that the donor star
is not a main sequence star, but the exposed core of a previously more
massive star. Post-supersoft X-ray binaries can, hence, be considered
failed SN\,Ia, and their observational hallmark should be abundances
typical of CNO burning. 

Testing the abundances of the donor star in a CV is equivalent to
testing the abundances of the material that it is transferring onto
the white dwarf, i.e. the abundances of the accretion disc/flow.  The
natural wavelength range to carry out this test is the far-ultraviolet
(FUV), which contains the N\,V\,$\lambda$\,1240 and
C\,IV\,$\lambda$\,1550 resonance lines. We are currently carrying out
a \textit{HST}/STIS FUV spectroscopic survey of a large number of CVs,
and G\"ansicke et al. (2003) presented the strong evidence that
EY\,Cyg, BZ\,UMa, EI\,Psc (1RXS\,J232953.9+062814) and V396\,Hya
(CE\,315) are post-supersoft X-ray binaries. Since the publication of
these results, our \textit{HST}/STIS survey has unveiled three
additional systems with significantly enhanced N/C emission line flux
ratios: QZ\,Ser, GS\,Pav, and CW1045+525. The fraction of likely
failed SN\,Ia in the \textit{HST}/STIS survey is $\sim12\%$, which is
within the range predicted by the evolutionary models. However,
considering the still rather small total number of systems observed in
this survey, the statistical significance of this fraction is too low
for a definite conclusion. Apart from the systems mentioned here,
AE\,Aqr, V1309\,Ori, TX\,Col, BY\,Cam, MN\,Hya, and GP\,Com have been
shown to exhibit strong N/C enhancement. The 14 post-supersoft CVs
known so far span a large range in orbital periods ($47-661$\,min) and
CV subclasses (dwarf novae, nova-like variables, polars, intermediate
polars, AM\,CVn systems). This underlines the fact that the observed
abundance anomalies are likely to be related to a general evolutionary
effect rather than to the specific properties of a small subgroup of
CVs.

\acknowledgements

I am grateful to all my friends and colleagues who have contributed to
this endeavour over the last years: Sofia Araujo-Betancor, Heinz
Barwig, Klaus Beuermann, Domitilla de Martino, Dieter Engels, Bob
Fried, Hans Hagen, Emilios Harlaftis, Ivan Hubeny, Christian Knigge,
Knox Long, Tom Marsh, Ronald Mennickent, Daisaku Nogami, Pablo
Rodr\'\i guez-Gil, Linda Schmidtobreick, Matthias Schreiber, Robert
Schwarz, Andreas Staude, Ed Sion, Paula Szkody, Claus Tappert, John
Thorstensen.

\end{document}